\documentstyle[12pt,epsfig,rotating]{article}
\def\pge{\pagestyle{empty}} \def\pgn{\pagestyle{plain}}
\def\spg{\setcounter{page}} 
\def\bd{
\begin{document}} \def\ed{\end{document}}
\def\bmp{\begin{minipage}} \def\emp{\end{minipage}}
\def\bcc{\begin{center}} \def\ecc{\end{center}}     \def\npg{\newpage}
\def\beq{\begin{equation}} \def\eeq{\end{equation}} \def\hph{\hphantom}
\def\be{\begin{equation}} \def\ee{\end{equation}} \def\r#1{$^{[#1]}$}
\def\n{\noindent} \def\ni{\noindent} \def\pa{\parindent} 
\def\hs{\hskip} \def\vs{\vskip} \def\hf{\hfill} \def\ej{\vfill\eject} 
\def\cl{\centerline} \def\ob{\obeylines}  \def\ls{\leftskip}
\def\underbar#1{$\setbox0=\hbox{#1} \dp0=1.5pt \mathsurround=0pt
   \underline{\box0}$}   \def\ub{\underbar}    \def\ul{\underline} 
\def\f{\left} \def\g{\right} \def\e{{\rm e}} \def\o{\over} 
\def\vf{\varphi} \def\pl{\partial} \def\cov{{\rm cov}} \def\ch{{\rm ch}}
\def\la{\langle} \def\ra{\rangle} \def\EE{e$^+$e$^-$}
\def\bitz{\begin{itemize}} \def\eitz{\end{itemize}}
\def\btbl{\begin{tabular}} \def\etbl{\end{tabular}}
\def\btbb{\begin{tabbing}} \def\etbb{\end{tabbing}}
\def\beqar{\begin{eqnarray}} \def\eeqar{\end{eqnarray}}
\def\\{\hfill\break} \def\dit{\item{-}} \def\i{\item} 
\def\bbb{\begin{thebibliography}{9} \itemsep=-1mm}
\def\ebb{\end{thebibliography}} \def\bb{\bibitem}
\def\bpic{\begin{picture}(260,240)} \def\epic{\end{picture}}
\def\akgt{\noindent{\bf Acknowledgements}}
\def\fgn{\noindent{\bf\large\bf Figure captions}}
%%%%%%%%%%%%%%%%%%%%%%%%%%%%%%%%%%%%%%%%%%%%%%%%%%%%%%%%%%%%%%%%%%%%
\bd
\pge
%%%%%%%%%%%%%%%%%%%%%%%%%%%%%%%%%%%%%%%%%%%%%%%%%%%%%%%%%%%%%%%%%%%%
\vskip-6.5cm
\hskip12cm{\large HZPP-9903}

\hskip12cm{\large March 10, 1999}

\vskip1.5cm

\begin{center}

{\Large The Influence of Statistical Fluctuations }

{\Large on Erraticity Behavior of Multiparticle System
\footnote{ \ This work is supported in part by the U. S. Department of Energy 
under Grant \\
\null{} \hs0.6cm No. DE-FG06-91ER40637 and the National 
Natural Science Foundation of China \\ 
\null{} \hs0.6cm (NSFC) under Grant No.19575021.}}
\vskip0.5cm

{Fu Jinghua $^\dag$  \ \ \ \ \ \ \ Wu Yuanfang $^\ddag$ $^\dag$ 
 \ \ \ \ \ \ \ Liu Lianshou $^\dag$}
\vskip0.0cm

{\small\dag \  Institute of Particle Physics, Huazhong Normal University, 
Wuhan 430079 China}
\vskip0.0cm

{\small\ddag \ Department of Physics, University of Oregon, Eugene 
OR 97403, USA}

\vskip0.0cm

{\small Tel: 027 87673313 \qquad FAX: 027 87662646 
\qquad email: liuls@iopp.ccnu.edu.cn}
\date{ }
\begin{minipage}{125mm}
\vskip 1.5cm
\begin{center}{\Large Abstract}\end{center}
\vskip 0.0cm
\ \ \ \  It is demonstrated that in low multiplicity sample, 
the increase of the fluctuation of event-factorial-moments 
with the diminishing of phase space scale, called ``erraticity'',
are dominated by the statistical fluctuations. 
The erraticity behavior observed at NA27 experiment
can be readily reproduced by pure statistical fluctuations.   
Applying erraticity analysis to a high multiplicity sample  
is recommended and the method is improved at very high multiplicity
case as well.

\end{minipage}
\end{center}
\vskip 1in
{\large PACS number: 13.85 Hd
\vskip0.2cm

\ni
}

\npg \pgn \spg{2}

Since the finding of unexpectedly large local fluctuations 
in a high multiplicity event recorded by the JACEE 
collaboration~\cite{JACEE}, the investigation of non-linear phenomena in 
high energy collisions has attracted much attention~\cite{Kittel}. The
anomalous scaling of factorial moments, defined as 
\beqar %%(1)
F_q &=& 
    \frac{1}{M}\sum\limits_{m=1}^{M}
    \frac{\la n_m(n_m-1) \cdots (n_m-q+1)\ra} 
    {\la n_m \ra ^q  } ,  
\eeqar
at diminishing phase space scale or increasing division number 
$M$ of phase space~\cite{BP}:
\beqar %%(2)
F_q \propto M^{-\phi_q},
\eeqar
called intermittency (or fractal) has been proposed for this purpose
in multiparticle system. The average $\la \cdots \ra$ in Eqn.(1) is over 
the whole event sample and $n_m$ is the number of particle falling in the
$m$th bin.  That kind of anomalous scaling has been observed successfully
in various experiments~\cite{NA22}\cite{NA27}.

A recent new development further along that direction is the event-by-event
analysis~ \cite{BiaZiaja}\cite{Liu}. An important step in this kind of 
analysis
was made by  Cao and  Hwa~\cite{CaoHwa}, who first pointed out the 
difference between a dynamic system in which time sequence can be traced 
and the multiparticle system where only (phase) space patterns can be 
obtained. They proposed to measure the pattern by the event factorial 
moments
\beqar  %%(3)
F_q^{({\rm e})} &=& \frac
    { \frac{1}{M}\sum\limits_{m=1}^{M} n_m(n_m-1) \cdots (n_m-q+1)} 
    { \f( \frac{1}{M}\sum\limits_{m=1}^{M} n_m\g)^q}
\eeqar
as opposed to sample-factorial-moments 
defined in Eqn.(1) averaged over all events. 
Its fluctuations from event to event can be quantified
by its normalized moments as:
\begin{equation} %%(4) 
C_{p,q}=\la \Phi_q^p\ra, \quad \Phi_q= {F_q^{(e)}} \f/ 
\la F_q^{(e)}\ra \g.,
\end{equation}
\noindent and by $dC_{p,q}/dp$ at $p=1$:
\beqar   %%% (5)
\Sigma_q=\la \Phi_q \ln \Phi_q \ra
\eeqar
If there is a power law behavior of the fluctuation as 
division number goes to infinity, or as resolution $\delta=\Delta/M$ 
goes to very small, {\rm i.e.},
\begin{equation}   %%% (6)
C_{p,q}(M) \propto M^{\psi_q(p)},
\end{equation}
the phenomenon is referred to as erraticity~\cite{hwa}.
The derivative of exponent $\psi_q(p)$ at $p=1$ 
\begin{equation}   %%% (7)
\mu_q=\f.\frac{d}{dp}\psi_q(p)\g|_{p=1} = 
             \frac{\partial \Sigma_q}{\partial \ln M}.
\end{equation}
describes the width of the fluctuation and so is called as 
entropy index. In the following, we will 
call $C_{p,q}$ or $\Sigma_q$ in Eqn.(4) and (5) as erraticity-moments. 

It is well known that the obstacle of event-by-event analysis is  
the influence of statistical fluctuations caused by insufficient number
of particles. The big advantage of sample factorial moments in
Eqn.(1) is that it can eliminate this kind of statistical fluctuations. 
It has been proved~\cite{BP} that if the statistical fluctuations of 
particles falling in a bin is Possonian like, then the sample-factorial-moments 
equal to the corresponding dynamic probability moments:
\beqar %%(8)
F_q=C_q=M^{q-1}\sum\limits_{1=1}^{M} \la p_i^q \ra .
\eeqar
\noindent Again, the average is over the whole sample. The $p_i$ is the
probability of particle falling in the $i$th bin in a certain event. 
However, we can not follow the same procedure to get the similar equation
for event factorial moments of Eqn.(3). Since the number of 
particles in an event is not large enough and so is the number of bin,  event 
factorial moments can not completely eliminate statistical fluctuations
and present the dynamic probability
moments associated with it. How large of the statistical 
fluctuations in erraticity 
analysis is and how it depends on the number of multiplicity have not
been seriously estimated yet. We will answer these
questions quantitatively in the letter. 

To be direct and obvious, we firstly use an unique flat probability 
distribution in whole studying interval and whole sample. It means
that the probabilities in all bins are equal and are the same for 
different events. For simplicity, we use only the fixed 
number of multiplicity. In this case the 
denominator in the definition of factorial moment, eqn.(1) and (3),
becomes simply $N(N-1)$~\cite{BP}. We first take $N=9$, which is about the 
average multiplicity at ISR energies. The distribution of particle 
in the whole studying phase region
in an event can be readily located by Bernouli distribution:
\beq   %%% (9)
B(n_1,\dots,n_M|p_1,\dots,p_M) = \frac{N!}{n_1!\cdots n_M!} 
p_1^{n_1}\cdots p_M^{n_M} , \qquad \sum_{m=1}^M n_m=N.
\eeq
By this way, we simulate a sample with 1000 events. The results of 
the second order sample-factorial-moments $F_2$, erraticity-moments $C_{p,2}$ 
and $\Sigma_2$ on the division number $M$ of the phase-space region
are shown in Fig.1($a$).  The second order sample-factorial-moment is a 
constant with the increasing of division number. This is what we expect.
Since no dynamic is input, it becomes a constant after eliminating
the statistical fluctuations. While the increase of erraticity-moments 
$C_{p,2}$ and $\Sigma_2$ with division number is measurably
large. These contributions come from pure statistical fluctuations of
event factorial moments due to the insufficient number of particle in 
an event since there is no dynamic fluctuation from event to event in the
case. This results can fully recover what has 
observed in NA27 data~\cite{WSS}, cf. the open circles in the 
second figure of Fig.1($a$).  This means that in low multiplicity
sample, the statistical fluctuations of event factorial moments 
dominate the erraticity behavior of multiparticle system. Event factorial 
moments is not a good representation of event dynamic at low multiplicity
events.

However, erraticity analysis proposed a very important way to study the
event-by-event fluctuations though we are still not clear whether there
are such fluctuations and if there are what mechanism
causes them. We have demonstrated in our former paper~\cite{FWL} 
that if and only if different events have different dynamic fluctuation
strengths, the erraticity moments will keep increasing with the increasing 
of division number and so has nonzero entropy index.  

As is well known, the statistical fluctuations will become negligible
if the multiplicity of an event is large enough. At how high a multiplicity
the event factorial moments can measure the dynamic fluctuations of
a finite particle
system is a very meaningful question. In the left of the
paper, we will focus our discussion on answering the question.

Now we switch the fixed multiplicity $N$ to 20 and 300 
in the above mentioned simulation. The corresponding  second order 
sample-factorial-moments $F_2$, erraticity-moments $C_{p,2}$ 
and $\Sigma_2$ versus the division number $M$ of the phase-space region
are shown in Fig.1($b$) and ($c$) respectively.  
The second order sample-factorial-moment keep to be a constant as we know.
The erraticity moments become flater and flater with the increase of 
multiplicity. It means that pure statistic fluctuations of event factorial 
moments are greatly depressed by the increase of multiplicity. 

%%% >From Fig.(1), we can see that, when multiplicity is larger than $300$, 
From Fig.(1), we can see that, when multiplicity is larger than $300$, 
event factorial moments can be
approximately used to describe the event spatial pattern associated with 
it and its moments --- erraticity moments --- can represent the
erraticity behavior of the system safely.

In order to confirm this upper limit of multiplicity, we do following
parallel analysis for a system with dynamic fluctuation from event to event. 
As we know~\cite{FWL},  
random cascading model, or $\alpha$-model is the simplest model which
can be used to generate a sample with nonzero entropy index. 
We will use it for our quantitative discussion below.
In the random cascading $\alpha$-model, the $M$ division of 
a phase space region $\Delta$ is made in steps. At the first step, 
it is divided into two equal parts; at the second step,  
each part in the first step is further divided into two equal parts, 
and so on. The steps are repeated until $M= {\Delta Y / \delta y}=2^{\nu}.$
How particles are distributed from step-to-step between the two
parts of a given phase space cell is defined by the independent random variable 
$ \omega_{\nu j_{\nu}}$, where $j_{\nu}$ is the position of the sub-cell 
($1\le j_{\nu}\le 2^{\nu}$) and $\nu$ is the number of steps.
It is given by~\cite{KXTB}:
\beq  %%% (10)
\omega_{\nu,2j-1}={1\over 2}(1+\alpha r) \ \ \ ; \ \ \ 
 \omega_{\nu,2j}={1\over 2}(1-\alpha r), \qquad j=1,\dots,2^{\nu-1}
\eeq
where, $r$ is a random number distributed uniformly in the interval
$[-1,1]$. $\alpha$ is a positive number less than unity, which 
determines the region of the random variable $\omega$ and describes 
the strength of dynamical fluctuations in the model. 
If it change from event to event, there will be different dynamic fluctuation
strength in different events. Here, let 
it has a Gaussian distribution. 
The mean and variance of the Gaussian are both chosen as 0.22.
After $\nu$ steps,
the probability in the $m$th window ($m=1,\dots,M$) is 
$p_m=\omega_{1j_1}\omega_{2j_2}\dots \omega_{\nu j_{\nu}}$. 

By the model, we generate 1000 events. The intermittency analysis, or
the logarithm of second order sample probability moment ln$C_2$, 
and erraticity moments 
$\ln C_{p,2}$ and $\Sigma_2$ as function of $\ln M$ are shown in 
Fig.2($a$). Now the sample probability moment ln$C_2$ has a power law 
behavior as dynamic fluctuations have been input. The erraticity moments 
also show a power law behavior at large division number region. It
represents the dynamic fluctuation from event to event. The corresponding
entropy index obtained from a linear fit to the last 3 points
of $\Sigma_2$ is $\mu_2=0.0161$. 

Finite number of particle can also be added to the above pure dynamic 
fluctuation model by Bernouli distribution of Eqn.(9).  Again, we 
put $N=9$ first. The corresponding factorial moment and erraticity moments
are given in Fig.2($c$).  The value of erraticity moments 
now are much larger than those obtained from the original 
pure dynamic fluctuations in Fig.2($a$). 
The entropy index, $\mu_2=0.422$, also turns out to be more than one magnitude 
bigger. This results confirm us again that the erraticity behavior
is dominated by statistical fluctuations in low multiplicity sample if we
use event factorial moments to characterize it. 
Though there is dynamic fluctuation from event to event, it will be merged 
to large statistical fluctuations in the case. 

Secondly, we let $N=300$. The corresponding factorial and erraticity moments 
are shown in Fig.2($b$). The erraticity moments now approach to its original
dynamic fluctuation values in Fig.2($a$) and entropy index is $\mu_2=0.0168$ 
close to its real value $0.0161$.   So we get the same conclusion 
as pure statistic fluctuation case. After multiplicity is larger than $300$, 
erraticity behavior of the system can be estimated by the fluctuation
of event factorial moments.

To show quantitatively the influences of statistic fluctuations 
on erraticity behvior at different multiplicity cases,
we simulate various number $N=5, \dots, 1000$ of particles in an event 
for both flat probability distribution and above-described $\alpha$ model 
cases. The corresponding entropy indices 
are given in Fig.3 as full circles (flat distribution) and full triangles 
($\alpha$ model) respectively. We can see 
that both of them decrease with multiplicity. For flat probability 
distribution, entropy index of statistical fluctuation is depressed more 
than three orders of magnitude when multiplicity $N$ increases from a few 
to $300$. After multiplicity $N > 300$, the entropy index is unmeasurably 
small.  Meanwhile, the entropy index of $\alpha$-model sample approaches 
to its real dynamic value $\mu_2=0.0161$, cf. the solid line in Fig.2,
after $N > 300$.   
The multiplicity of current and nearly coming heavy-ion collision is
about this number or higher. The erraticity analysis given by event 
factorial moments is
applicable for heavy-ion collisions, when the average multiplicity 
is higher than $300$, where we are free from the influence of the 
statistical fluctuations. 

In fact, if multiplicity is higher than a thousand, which has been
recorded in NA49 experiments and will be the case for the future
heavy-ion collision experiments, the 
factorial moments analysis of a single event is unnecessary anymore as 
$n_m(n_m-1)\cdots$ does not make much difference from 
$n_m\cdot n_m\cdots$ in most of the phase-space
bins which provides main contribution to the anomalous scaling of moments. 
In these cases the probability distribution in an 
event can be approximately presented by:
\beqar %%  (11)
p_m \approx \frac {n_m}{N}, \quad 1=\sum\limits_{m=1}^M p_m .
\eeqar
The erraticity-moments of event-probability-moments can be 
consequently defined by:
\beqar  %%% (12)
C_{p,q} = \la \Phi_q^p \ra, \quad 
\Phi_q = {\sum_{m=1}^M p_m^q } \f/ \la {\sum_{m=1}^M p_m^q} \ra \g.
\eeqar
By this definition, we repeat the analysis for both the flat 
probability distribution and the dynamic-fluctuation distribution cases.
It is a little bit smaller than the corresponding
event-factorial-moments analysis at flat probability distribution case,
cf. the open and full circles in Fig .3,
and so it depresses the influence of statistical fluctuations more.
The dynamic fluctuations of event to event in the $\alpha$ model case
can still be abstracted out as done by the event-factorial-moments
description, cf. the open and full triangles in Fig.3.

%%% >From the simple discussion above, we can make the following conclusions:
From the simple discussion above, we can make the following conclusions:
If we use event factorial moments to measure spatial pattern, in very 
low multiplicity sample, such as the sample of ISR energies,
statistical fluctuations caused by insufficient number of particle in
an event will control the erraticity behavior of the system. Therefore,
the physical
conclusions from the experimental data on these kind of sample can not be
treated seriously. However, if the multiplicity of studying
sample is larger than $300$, the influence of statistical fluctuations
on erraticity behavior is negligible. Therefore, the erraticity behavior,
if any, could be well observed in current and future heavy-ion collisions.
Further more, if the multiplicity of an event is larger than a thousand, 
the probability 
moments defined by Eqn.(11)(12) can present erraticity behavior of the system
as well as the event-factorial-moments.

\akgt

One of the authors (Wu Yuanfang) would like to thank Prof. Rudolph C. Hwa for
leading her to the field and many helpful suggestions and comments on 
the paper.

\newpage
%%%%%%%%%%%%%%%%%%%%%%%%%%%%%%%%%%%%%%%%%%%%%%%%%%%%%%%%%%
\def\Journal#1#2#3#4{{#1} {\bf #2} (#3) #4}
\def\NCA{\em Nuovo Cimento} \def\NIM{\em Nucl. Instrum. Methods}
\def\NIMA{{\em Nucl. Instrum. Methods} A} \def\NPB{{\em Nucl. Phys.} B}
\def\PLB{{\em Phys. Lett.}  B} \def\PRL{\em Phys. Rev. Lett.}
\def\PRD{{\em Phys. Rev.} D} \def\ZPC{{\em Z. Phys.} C}
\def\PRE{{\em Phys. Rev.} E} \def\PRC{{\em Phys. Rev.} C} 
%%%%%%%%%%%%%%%%%%%%%%%%%%%%%%%%%%%%%%%%%%%%%%%%%%%%%%%%%%

\newpage

\ni{\Large\bf Figure Captions}
\vs1cm

{\pa=0pt{\ls=15mm\rightskip15mm
\hs-15mm
{\bf Fig.1} \ ($a$) The dependence of the logarithm of the
second order sample-factorial-moments $F_2$, erraticity-moments $C_{p,2}$ 
and $\Sigma_2$ on that of the phase-space division number $M$ for a flat
probability distribution with particle number equal to 9 ($a$), 20 ($b$)
and 300 ($c$) respectively.   The dashed lines are linear fit. Open points
are experimental results of NA27. The solid points are MC results. The solid
lines are for guiding the eye.
\par}}

\vs1cm
{\pa=0pt{\ls=15mm\rightskip15mm
\hs-15mm
{\bf Fig.2} \ ($a$) The dependence of the logarithm of the
second order probability-moments $C_2$,
erraticity-moments $C_{p,2}$ and $\Sigma_2$
on that of the phase-space division number $M$ for the $\alpha$
model with Gaussian-distributed $\alpha$. ($b$) The same as ($a$) but for
the sample-factorial-moments $F_2$  
and the corrssponding erraticity-moments $C_{p,2}$ and $\Sigma_2$
with 300 particles. (c) The same as ($b$) but
with 9 particles.  The dashed lines are linear fit. The solid
lines are for guiding the eye.
\par}}

\vs1cm
{\pa=0pt{\ls=15mm\rightskip15mm
\hs-15mm
{\bf Fig.3} \ The dependence on number of particle of the entropy
indices $\mu_2$ for Gaussian $\alpha$-model calculated from
event-factorial-moments (full triangles) and from probability-moments
(open triangles). The same for flat distribution (full and open circles).
The solid line is the dynamical result without statistical fluctuation.
The dashed lines are for guiding the eye.
\par}}
\newpage
\baselineskip 0.18in

\begin{picture} (260,240) 
\put(-95,-400)   
{\epsfig{file=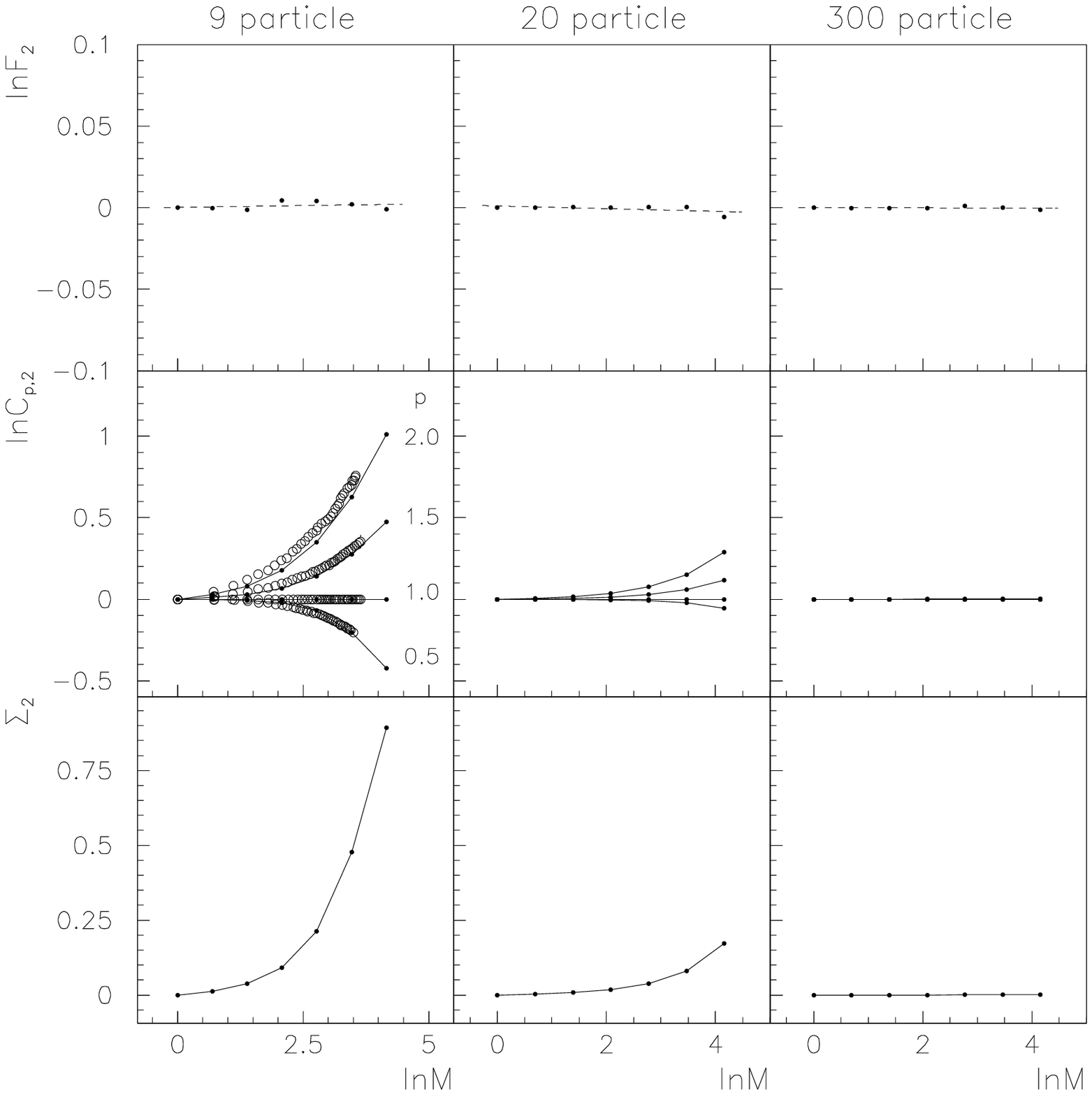,bbllx=0cm,bblly=0cm,
           bburx=8cm,bbury=6cm}}  
\end{picture}
 \vs9.0cm
\cl{($a$)\hs5cm ($b$) \hs5cm ($c$)}
\vs1.2cm
{\pa=0pt{\ls=15mm\rightskip15mm
\hs-15mm
{\bf Fig.1} \ ($a$) The dependence of the logarithm of the
second order sample-factorial-moments $F_2$, erraticity-moments $C_{p,2}$ 
and $\Sigma_2$ on that of the phase-space division number $M$ for a flat
probability distribution with particle number equal to 9 ($a$), 20 ($b$)
and 300 ($c$) respectively.   The dashed lines are linear fit. The solid
lines are for guiding the eye.
\par}}

\newpage

\begin{picture} (260,240) 
\put(-75,-400)   
{\epsfig{file=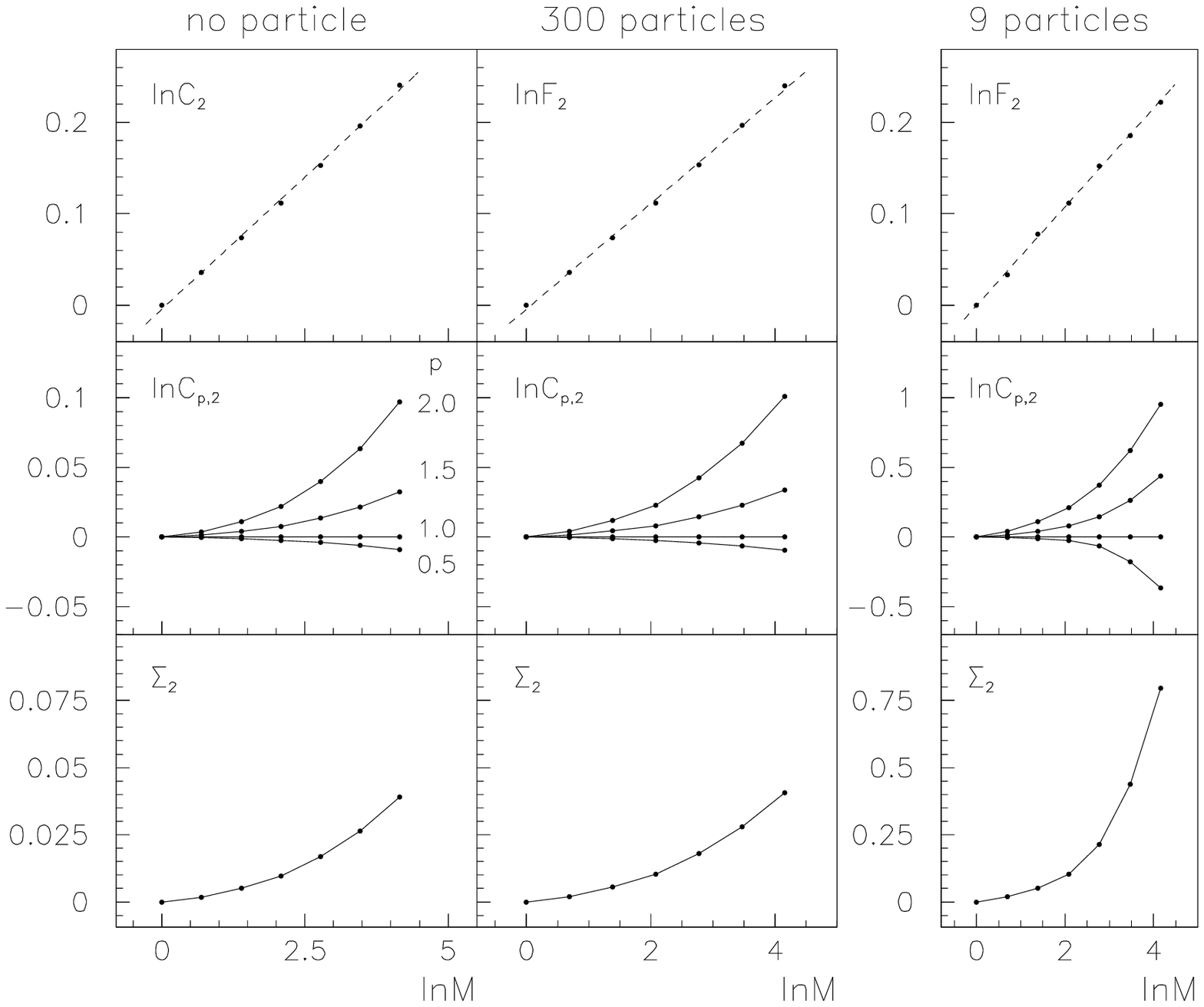,bbllx=0cm,bblly=0cm,
           bburx=8cm,bbury=6cm}}  
\end{picture}
\vs7.5cm
\hs3.0cm{($a$)\hs4.5cm ($b$) \hs5.5cm ($c$)}
\vs1cm

{\pa=0pt{\ls=35mm\rightskip15mm
\hs-15mm
{\bf Fig.2} \ ($a$) The dependence of the logarithm of the
second order probability-moments $C_2$,
erraticity-moments $C_{p,2}$ and $\Sigma_2$
on that of the phase-space division number $M$ for the $\alpha$
model with Gaussian-distributed $\alpha$. ($b$) The same as ($a$) but for
the sample-factorial-moments $F_2$  
and the corrssponding erraticity-moments $C_{p,2}$ and $\Sigma_2$
with 300 particles. (c) The same as ($b$) but
with 9 particles.  The dashed lines are linear fit. The solid
lines are for guiding the eye.
\par}}

\newpage

\begin{picture} (260,240) 
\put(-75,-420)   
{\epsfig{file=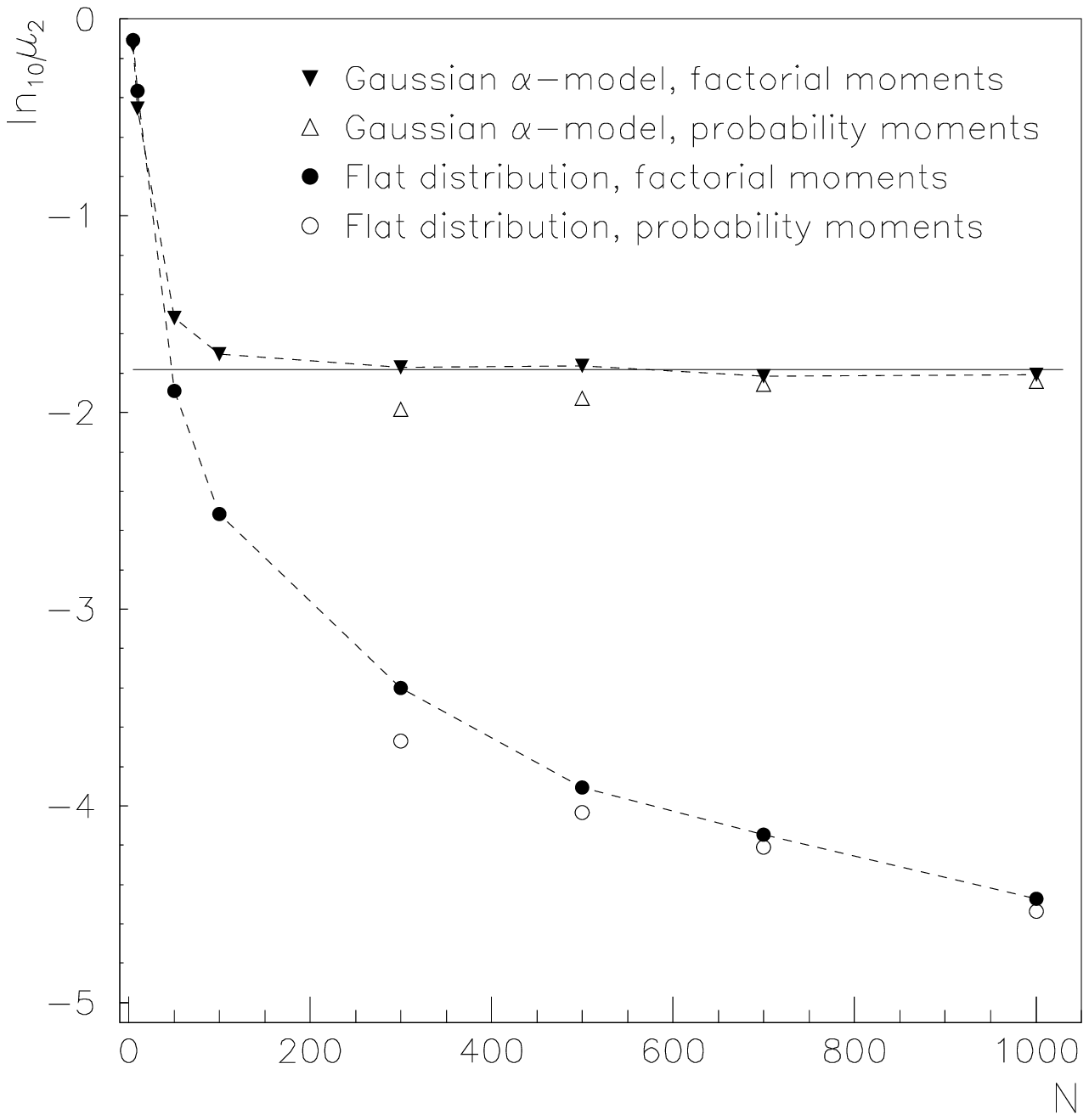,bbllx=0cm,bblly=0cm,
           bburx=8cm,bbury=6cm}}  
\end{picture}
\vskip10.0cm

{\pa=0pt{\ls=35mm\rightskip15mm
\hs-15mm
{\bf Fig.3} \ The dependence on number of particle of the entropy
indices $\mu_2$ for Gaussian $\alpha$-model calculated from
event-factorial-moments (full triangles) and from probability-moments
(open triangles). The same for flat distribution (full and open circles).
The solid line is the dynamical result without statistical fluctuation.
The dashed lines are for guiding the eye.
\par}}
\ed